\documentclass{article}
\usepackage[utf8]{inputenc}
\usepackage{geometry}
\usepackage{graphicx}
\usepackage[usenames,dvipsnames]{xcolor}
\usepackage{bm}
\usepackage{amsmath}
\usepackage[normalem]{ulem}
\usepackage{graphicx}
\usepackage{color,soul}
\usepackage[colorlinks=true, citecolor=black, urlcolor=blue]{hyperref}
\usepackage{authblk}
\usepackage[normalem]{ulem}

\begin{document}

\title{Screening of Therapeutic Agents for COVID-19 using Machine Learning and Ensemble Docking Simulations}

\author[1]{Rohit Batra}
\author[1,2]{Henry Chan}
\author[3]{Ganesh Kamath}
\author[4]{Rampi Ramprasad}
\author[1]{Mathew J. Cherukara}
\author[1,2]{Subramanian Sankaranarayanan}
\affil[1]{Center for Nanoscale Materials, Argonne National Laboratory, Lemont, Illinois 60439, United States}
\affil[2]{Department of Mechanical and Industrial Engineering, University of Illinois,Chicago, Illinois 60607, United States}
\affil[3]{Dalzielfiver LLC, 3500 Carlfield St., El Sobrante, CA 94803, United States}
\affil[4]{School of Materials Science and Engineering, Georgia Institute of Technology, 771 Ferst Drive NW, Atlanta, Georgia 30332, United States}

\date{\today}
\maketitle
\begin{abstract}
The world has witnessed unprecedented human and economic loss from the COVID-19 disease, caused by the novel coronavirus SARS-CoV-2. Extensive research is being conducted across the globe to identify therapeutic agents against the SARS-CoV-2. Here, we use a powerful and efficient computational strategy by combining machine learning (ML) based models and high-fidelity ensemble docking simulations to enable rapid screening of possible therapeutic molecules (or ligands). Our screening is based on the binding affinity to either the isolated SARS-CoV-2 S-protein at its host receptor region or to the Sprotein-human ACE2 interface complex, thereby potentially limiting and/or disrupting the host-virus interactions. We first apply our screening strategy to two drug datasets (CureFFI and DrugCentral) to identify hundreds of ligands that bind strongly to the aforementioned two systems. Candidate ligands were then validated by all atom docking simulations. The validated ML models were subsequently used to screen a large bio-molecule dataset (with nearly a million entries) to provide a rank-ordered list of $\sim$19,000 potentially useful compounds for further validation. Overall, this work not only expands our knowledge of small-molecule treatment against COVID-19, but also provides an efficient pathway to perform high-throughput computational drug screening by combining quick ML surrogate models with expensive high-fidelity simulations, for accelerating the therapeutic cure of diseases.



\end{abstract}

\section{Introduction}
On March 11, 2020 World health organization (WHO) declared the novel coronavirus disease, COVID-19, as a pandemic. More than a million people across 203 countries have already been affected by this disease, with more than 50,000 lives lost globally (as of April 2, 2020). In addition, daily lives of millions of people have been impacted because of the mandatory lock-downs observed across the world, let alone the economic cost of this adversity. The COVID-19 disease is caused by a new coronavirus SARS-CoV-2, belonging to the SARS family (SARS-CoV). SARS-CoV-2 has already been sequenced and several studies focused on understanding its interaction with the human cells (or receptors) are ongoing \cite{wu2020new, yu2020decoding, tang2020origin, sun2020understanding, bai2020presumed, gralinski2020return,wang2020}. Screening of small-molecules or biomolecules with potential therapeutic ability against COVID-19 is also being conducted using theoretical and machine learning methods \cite{smith2020repurposing, nguyen2020potentially, xu2020nelfinavir, beck2020predicting, bung2020novo}.

Initial reports on SARS-CoV-2, and previous works on the general SARS coronavirus, have suggested close interactions between the viral spike protein (S-protein) of coronavirus with specific human host receptors, such as the Angiotensin-converting enzyme 2 (ACE2) receptor. It has been hypothesized that compounds that can reduce interactions between S-protein:ACE2 receptors could limit viral recognition of the host (human) cells and/or disrupt the host-virus interactions. To this end, Smith et al. \cite{smith2020repurposing} recently conducted virtual high-throughput screening of nearly 9000 small-molecules that bind strongly to either 1) the isolated S-protein of SARS-CoV-2 at its host receptor region (thus, hindering the viral recognition of the host cells) or 2) to the S-protein:human ACE2 receptor interface (thus, reducing the host-virus interactions). They successfully identified 77 ligands (24 of which have regulatory approval from the Food and Drug Administration, FDA, or similar agencies) that satisfied one of the above two criteria. Despite the vast chemical space (millions to billions of biomolecules) that can be potentially explored, they were severely limited by the number of candidate compounds (nearly 9000) that were considered in their work owing to the high computational cost of the ensemble docking simulations employed in their methodology.

Here, we build on the exemplary work of Smith et al. \cite{smith2020repurposing} and use their data set generated from autodocking/molecular modeling for training and validating machine learning (ML) models. This allows us to significantly expand the search space and screen millions of potential therapeutic agents against COVID-19. We use the binding affinities (or their Vina scores) of the ligand to the S-protein:ACE2 interface complex and the isolated S-protein as the screening criteria to identify promising candidates using our ML models. The ML models were applied to two drug datasets, namely, CureFFI and DrugCentral containing $\sim$1500 and $\sim$4000 ligands present in known drugs, respectively. We also deploy the ML model to rapidly screen through a huge BindingDB dataset that contains over millions of small bio-molecules, to identify candidates that bind strongly to either the isolated S-protein or the S-protein:human ACE2 receptor interface complex.

\begin{figure}
	\centering 
	\includegraphics[width=1\textwidth]{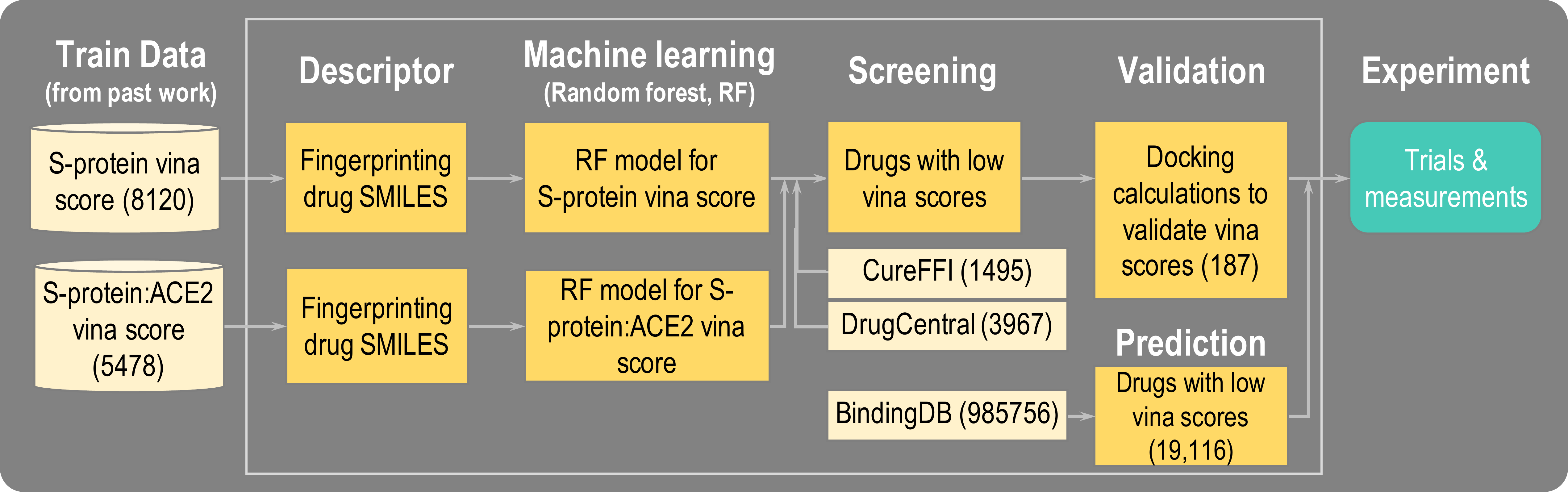}
	\caption{Overview of the workflow adopted to screen drug active ingredients with potential therapeutic capability for COVID-19. The white box denotes the stages performed in this work. The numbers within the bracket indicate the count of ligands in various datasets or stages of the workflow.}
	\label{fig:figure1}
\end{figure}

\begin{figure}
	\centering 
	\includegraphics[width=0.5\textwidth]{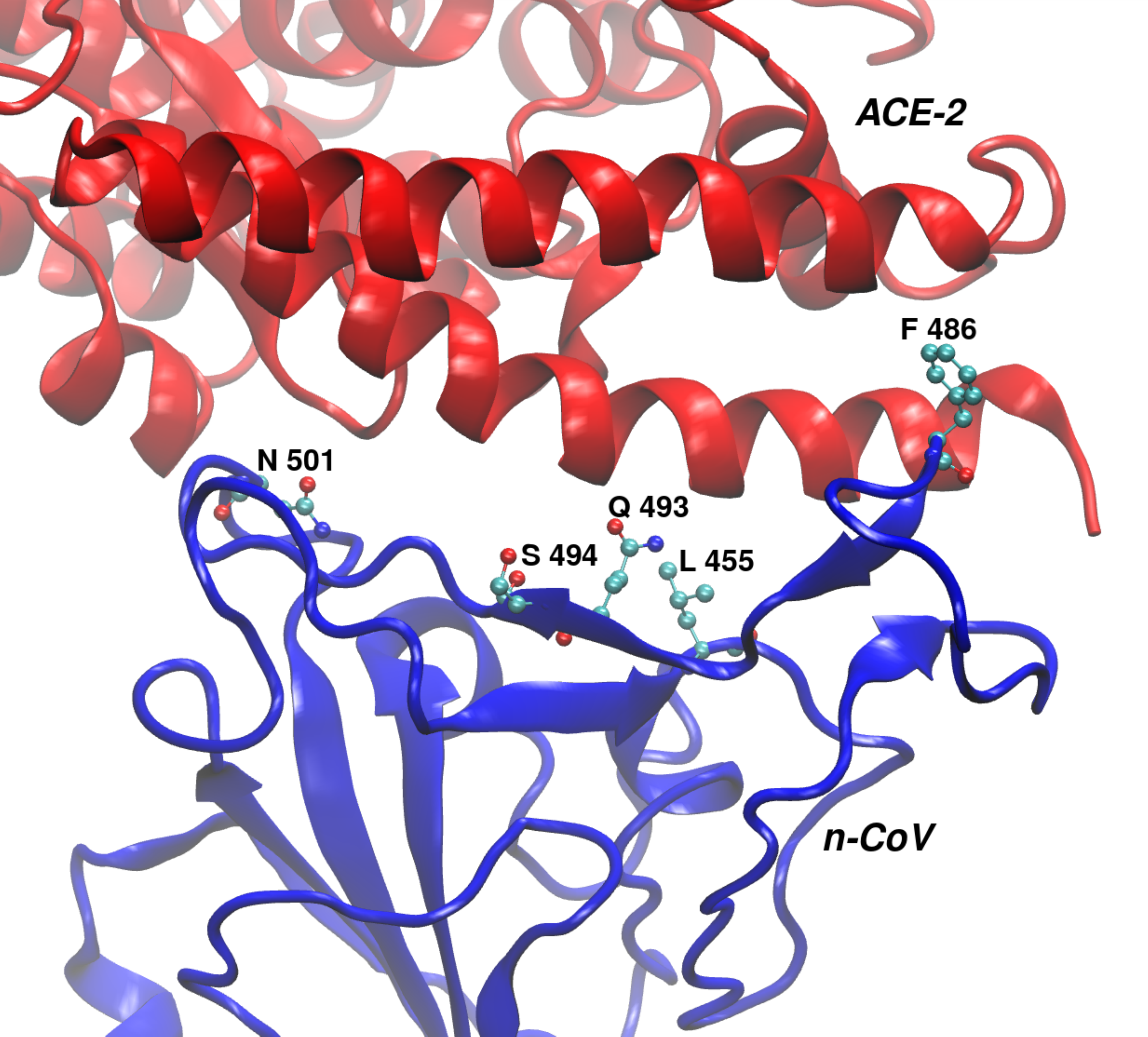}
	\caption{A representation of the interface between the coronavirus n-CoV (or SARS-CoV-2) and the human ACE2 receptor is shown, with the virus in blue and the human receptor in red. The mutations at the particular virus site are shown in CPK.}
	\label{fig:interface}
\end{figure}

The screening workflow adopted in this work 
is presented in Figure \ref{fig:figure1}, while an illustration of the interface between coronavirus SARS-CoV-2 and the ACE2 receptor in presented in Figure \ref{fig:interface}. Our work is based on combining ML methods with ensemble docking simulations to screen promising ligands.
Two independent random forest (RF) regression models were trained to quickly estimate the Vina scores of a given candidate drug molecule (or ligand) for the isolated S-protein and the S-protein:human ACE2 receptor interface using the datasets provided by Smith et al. \cite{smith2020repurposing}. The Vina score is a hybrid (empirical and knowledge-based) scoring function that ranks molecular conformations and predicts the free energy of binding based on inter-molecular contributions (\textit{e.g.}, steric, hydrophobic, and hydrogen bonding, etc.)\cite{trott2010autodock}. A set of hierarchical descriptors (or features/fingerprints) that capture different geometric and chemical information at multiple length-scales (atomic and morphological) were used to represent the molecules for successful application of the ML models. The models were validated by monitoring their performance on the validation set, and against ensemble docking simulations for hundreds of promising candidate ligands. Overall, we identified several hundreds of new ligand candidates with potential therapeutic ability against COVID-19, 75 of which are FDA approved. A list of $\sim$19,000 bio-molecules (from BindingDB dataset) satisfying the same screening criteria is also provided using the developed ML models. Based on the feature importance revealed by the ML models, we also provide analysis of key chemical trends that are common across the identified promising candidates.
We note that this work not only expands our knowledge of potential small-molecule treatment against COVID-19, but also provides a powerful and efficient pathway, i.e., training ML on results of computationally expensive simulations, using ML to cast a wider net, down-selection followed by targeted computational simulations, and finally chemical guidelines, for accelerating the therapeutic cure of other diseases.

\section{Methods}

\subsection{Machine learning}
As depicted in Figure \ref{fig:figure1}, starting from datasets of Vina scores from \cite{smith2020repurposing} for small molecules (for S-protein and S-protein:ACE2 interface), we first fingerprint the molecules using their SMILES representation; the fingerprinting procedure is discussed in detail below and in \cite{kim2018polymer}. The obtained molecular descriptors along with their respective Vina scores were then used to train two independent random forest ML models for each property. These ML models were next used to make Vina score predictions for the 1495 FDA approved drugs from CureFFI database, and 3967 other drugs from DrugCentral database. 187 candidate ligands with significantly low Vina score predictions (or high binding affinity) for both the S-protein and S-protein:ACE2 interface systems were screened for further validations using ensemble docking simulations. Our ML predictions were found to be in good agreement with the expensive docking simulations, thus, validating the developed ML model for its accurate Vina score predictions and identifying several tens of new FDA approved (or otherwise) ligands with high binding affinity to both the S-protein and S-protein:ACE2 interface. With the ML models validated, we apply them to an extensive small bio-molecule dataset, i.e., BindingDB with over millions of entries and screen many more potential candidates. Below we discuss in detail the training set, the molecular descriptors and the random forest algorithm used to develop the ML models.

\subsubsection{Training Dataset and Features}
Two training datasets were obtained from Smith et al. \cite{smith2020repurposing}, one corresponding to Vina score of a molecule with the S-protein and other for the S-protein:ACE2 interface complex. Each of the datasets contains 9127 molecules from the SWEETLEAD database \cite{novick2013sweetlead} along with their SMILES representations, which were used as input for our fingerprinting algorithm. For many molecules the Vina scores were reported to be extremely high (as much as 1,000,000 kcal/mol), while those with favorable binding energetics have Vina scores roughly in the -7 to 0 kcal/mol range. To remove such skewness in the data and train our models geared towards identifying favorable molecules with lower Vina scores, data points with only negative Vina scores were considered in this study. Further for a few cases, the SMILES representation could not be resolved correctly and were filtered out. Overall, this results in 5478 and 8120 data points (from the original number of 9127) for the S-protein:ACE2 interface and the isolated S-protein system, respectively. Henceforth, we refer to this cleaned dataset as the Smith dataset.

To build accurate and reliable ML models, it is important to include relevant features that collectively capture the trends in the Vina scores of different molecules towards S-protein and S-protein:ACE2 interface complex. The features should uniquely represent a molecule, and be readily available for new cases. Based on our past experience on fingerprinting organic materials including polymers\cite{kim2018polymer}, three hierarchical levels of features were considered capturing different geometric and chemical information about ligands at multiple length-scales. At the atomic scale, a count of a predefined set of motifs is included. The motifs are specified by the generic label ``A$_i$B$_j$C$_k$", representing an i-fold coordinated A atom, a j-fold coordinated B atom, and a k-fold coordinated C atom, connected in the specified order. For example, N3-C3-C4 represents a three-fold coordinated N, a three-fold coordinated carbon and a four-fold coordinated carbon \cite{huan2015accelerated}
At a slightly larger length-scale, quantitative structure-property relationship (QSPR) descriptors, often used in chemical and biological sciences, and implemented in the RDKit Python library, were used\cite{isarankura2009practical,nantasenamat2010advances,rdkit}. Lastly, at the highest length-scale, `morphological descriptors', such as length of the largest side-chain, shortest topological distance between rings, etc. were considered. More details on the different hierarchical descriptors can be found in our previous works \cite{kim2018polymer}.

\subsubsection{Machine learning Model}
The random forest (RF) regression algorithm, as implemented in the scikit-learn Python package\cite{scikitlearn}, was used to train the two Vina score models (S-protein and S-protein:ACE2 interface) using the Smith dataset. RF is an ensemble of decision trees, that averages predictions from a large group of `weak models' to overall result in a better prediction. It falls under the umbrella of ensemble methods, which are often the winning solutions in machine learning competitions.

The RF hyperparameters, i.e., the number of weak estimators, were estimated by maximizing the validation error during 5-fold cross-validation (CV), which leads to better generalization of the models and avoids overfitting. The performance of the ML models was evaluated using root mean square error (RMSE), mean absolute error (MAE) and correlation coefficient (R$^2$). To estimate prediction errors on unseen data, learning curves were generated by varying the sizes of the training and test sets, with results included in the supplementary information (SI). For learning curves, the test sets were obtained by excluding the training points from the Smith dataset. Additionally, for each random test-train split, statistically meaningful results were obtained by averaging over 10 runs. The final ML models used for prediction on the CureFFI, DrugCentral and BindingDB datasets were trained on the entire Smith dataset using 5-fold CV, and consisted of 400 and 700 estimators for the isolated S-protein and the S-protein:ACE2 interface datasets, respectively.

\subsection{Docking Computations}
To validate our ML models, we performed docking calculations of the top candidates identified by the models based on their low Vina scores. In-line with the works of Smith et al.\cite{smith2020repurposing}, we used the Autodock Vina software \cite{trott2010autodock} to compute binding affinities between the top candidates and the SARS-CoV-2 S-protein:ACE2 complex. The structure of the SARS-CoV-2 S-protein has a NCBI Reference Sequence YP\verb!_!009724390.1 and the ACE2 receptor has a protein data bank ID PDB:2AJF. Details regarding the construction and modeling of the SARS-CoV-2 S-protein:ACE2 complex are described here \cite{smith2020repurposing}. The SARS-CoV-2 S-protein has the necessary mutations from its predecessor SARS variety SARS-CoV, namely at L(455), F(486), Q(493), S (494), and N(501), respectively, which is illustrated in Figure \ref{fig:interface}. Smith et al. \cite{smith2020repurposing} focused on this binding pocket region and evaluated the binding affinities of different molecules from the SWEETLEAD library, as discussed earlier.

Following the procedure described by Smith et al. \cite{smith2020repurposing}, we also rank-ordered our top candidates based on their Vina scores, which is correlated to their free energy of binding to SARS-CoV-2 S-protein:ACE2. The docking receptors obtained from Smith et al. \cite{smith2020repurposing} consists of six conformations of SARS-CoV-2 S-protein:ACE2 interface as well as isolated SARS-CoV-2 S-protein receptor (\textit{i.e.}, with the ACE2 receptor removed), which were sampled using root mean squared displacement (RMSD) based clustering from 1.3 microsecond long all-atom Temperature Replica Exchange GROMACS simulations of the SARS-CoV-2 S-protein:ACE2 complex in water. The docking ligands were prepared from SMILES strings of the candidates using the Open Babel software \cite{o2011open}. Setup of the docking calculations is similar to that described by Smith et al. \cite{smith2020repurposing}, which defines a $1.2$ nm $\times$ $1.2$ nm $\times$ $1.2$ nm search space that encompasses the binding pocket located at the SARS-CoV-2 S-protein:ACE2 interface shown in Figure \ref{fig:interface}. The same search space was explored for the isolated SARS-CoV-2 S-protein receptor cases as well. For each candidate, the docking procedure finds the top $10$ optimized docking configurations and selects one with the best Vina score.
As described in more details below, the candidates identified by our ML model were obtained from the DrugCentral database as well as the FDA approved CureFFI database.


\subsection{Ligand Datasets}
While the Smith dataset\cite{smith2020repurposing} was used to train and validate the ML models, three additional drug datasets were used to make predictions and identify ligand candidates that show high binding affinity to the viral S-protein or the S-protein:ACE2 interface. These include 1) an all FDA approved CureFFI dataset \cite{cureffi}, 2) a dataset of common active ingredients from DrugCentral \cite{ursu2016drugcentral}, and a BindingDB dataset \cite{gilson2016bindingdb} of small molecules. SMILES representation of molecules were obtained from each of these datasets, and with some unprocessed candidates removed, resulted in 1495, 3967 and 985,756 entries, respectively. The CureFFI datasets consists of ligands approved by FDA and specifically contains central nervous system drugs. The drug list was parsed from an EPA (Environmental Protection Agency) suite and appropriately curated.
DrugCentral
is an open-access online drug compendium. It integrates the structure, bioactivity, regulatory, pharmacologic actions and indications for active pharmaceutical ingredients approved by FDA and other regulatory agencies. The BindingDB is a database publicly accessible based on measured binding affinities of drug-like molecules interacting with various protein targets and consists of more than a million entries of binding data and molecule datasets.
The first two datasets were exclusively used to validate the ML models against docking simulations, while the BindingDB dataset was used only for ML predictions.

\section{Results and Discussion}
\subsection{ML based Screening of FDA Approved and Other Ligands}
\begin{figure}
	\centering 
	\includegraphics[width=0.6\textwidth]{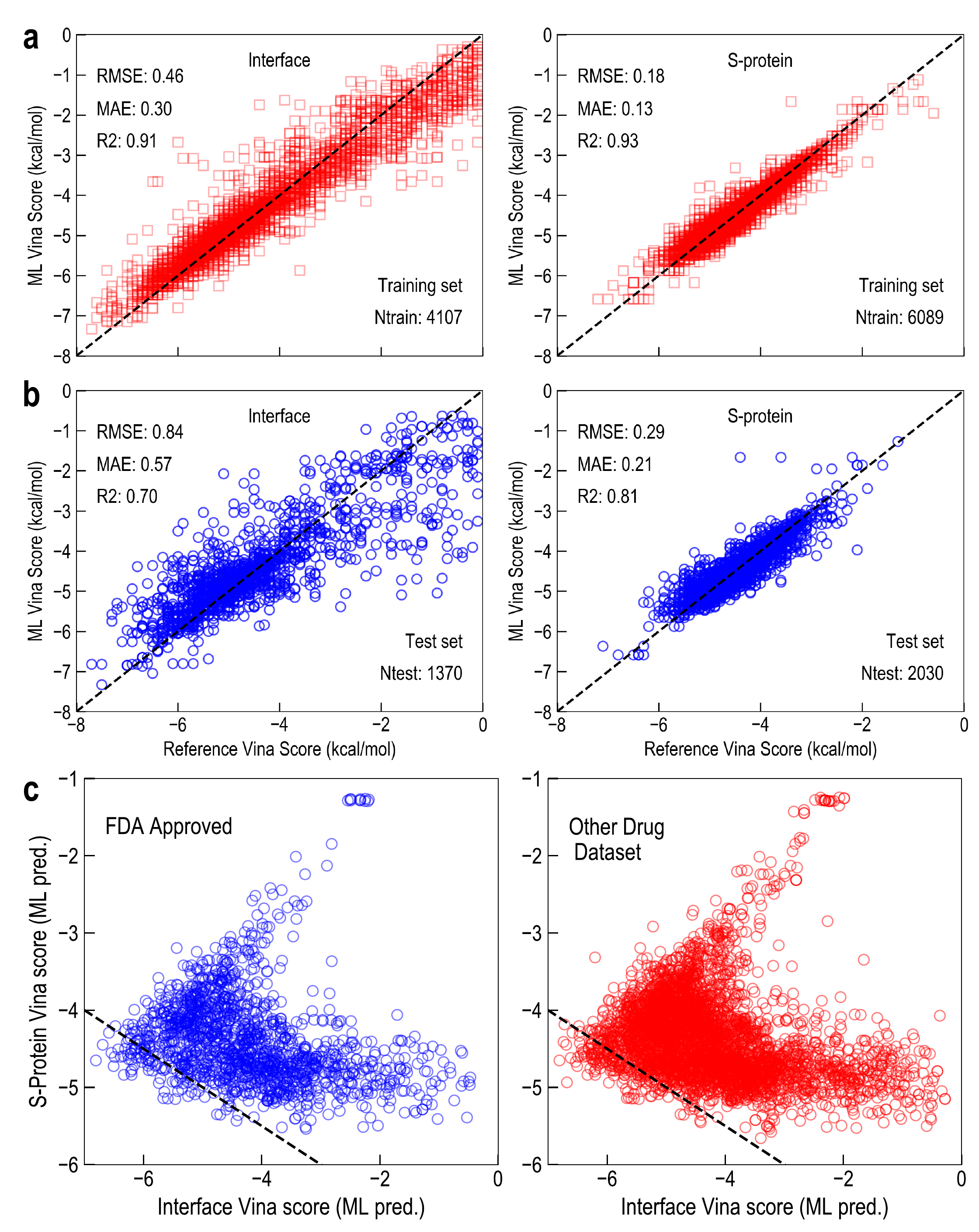}
	\caption{Parity plot of the S-protein and interface ML models for the (a) training and the (b) test set, demonstrating the good prediction accuracy achieved by both the ML models. (c) ML predictions of Vina scores for the isolated S-protein and S-protein:ACE2 receptor interface for FDA approved (left panel) and other drug (right panel) candidates obtained from CureFFI and DrugCentral databases. Candidate predictions below the dashed line were selected for further validation using docking simulations.}
	\label{fig:figure2}
\end{figure}

Figure \ref{fig:figure2}(a) and (b) present the performance results of the S-protein and S-protein:ACE2 interface RF models for the case when 75\% of Smith's dataset was used for training (with 5-fold CV), and the remaining 25\% as test set. The overall model performance on the test set is a good indicator of the expected errors on new candidate drugs with unknown Vina scores. Both the models can be seen to have good performance on the test set---a MAE of 0.21 kcal/mol was achieved for the S-protein model, while the S-protein:ACE2 model was only marginally worse with a MAE of 0.57 kcal/mol. Both these errors are well within typical chemical accuracy of 1 kcal/mol, and we believe the ML models are acceptable for screening purposes. Even for the S-protein:ACE2 model, relatively lower errors are observed for cases with low Vina scores, which are particularly more relevant to this study. See SI for more detailed validation of the ML models using learning curves, including error convergence studies on the training and test sets.

These results clearly indicate that the developed surrogate ML models could be used to quickly screen new ligand candidates with low S-protein or S-protein:ACE2 interface Vina scores without exclusively performing computationally demanding docking simulations. To this end, we use the ML models to make predictions for the FDA approved active ingredients in the CureFFI dataset and other ligands from the DrugCentral dataset, presented in Figure \ref{fig:figure2}(c). Since, the true Vina scores of these ligands is not known, here we only show their ML predictions. It has been hypothesized that a ligand could be effective against coronavirus if it either form S-protein:ACE2 interface-ligand binding complexes (low S-protein:ACE2 Vina score) to disrupt the host-virus interaction, or it binds to the receptor recognition region of the S-protein (low S-protein Vina score) to reduce viral recognition of the host. Thus, we define a simple screening criteria to select top candidates having low Vina scores on both the accounts. The dashed line in Figure \ref{fig:figure2}(c) depict the chosen screening criteria (given by equation $y = -\frac{x}{2} - 7.5$ with $x$, $y$ representing Vina scores for the S-protein:ACE2 interface-ligand complex and the S-protein-ligand system, respectively). All candidates having scores lower than this are screened for further validations. We note that 187 ligands were selected, from which 80 are FDA approved (CureFFI dataset), 107 are other drugs (DrugCentral dataset), and 29 are common to the Smith dataset. List of all 187 drugs (including their generic name and SMILES representation) and their Vina score predictions are provided in the SI.

\subsection{Validation using Docking Computations on FDA Approved and Other Ligands}
\begin{figure}
	\centering 
	\includegraphics[width=0.65\textwidth]{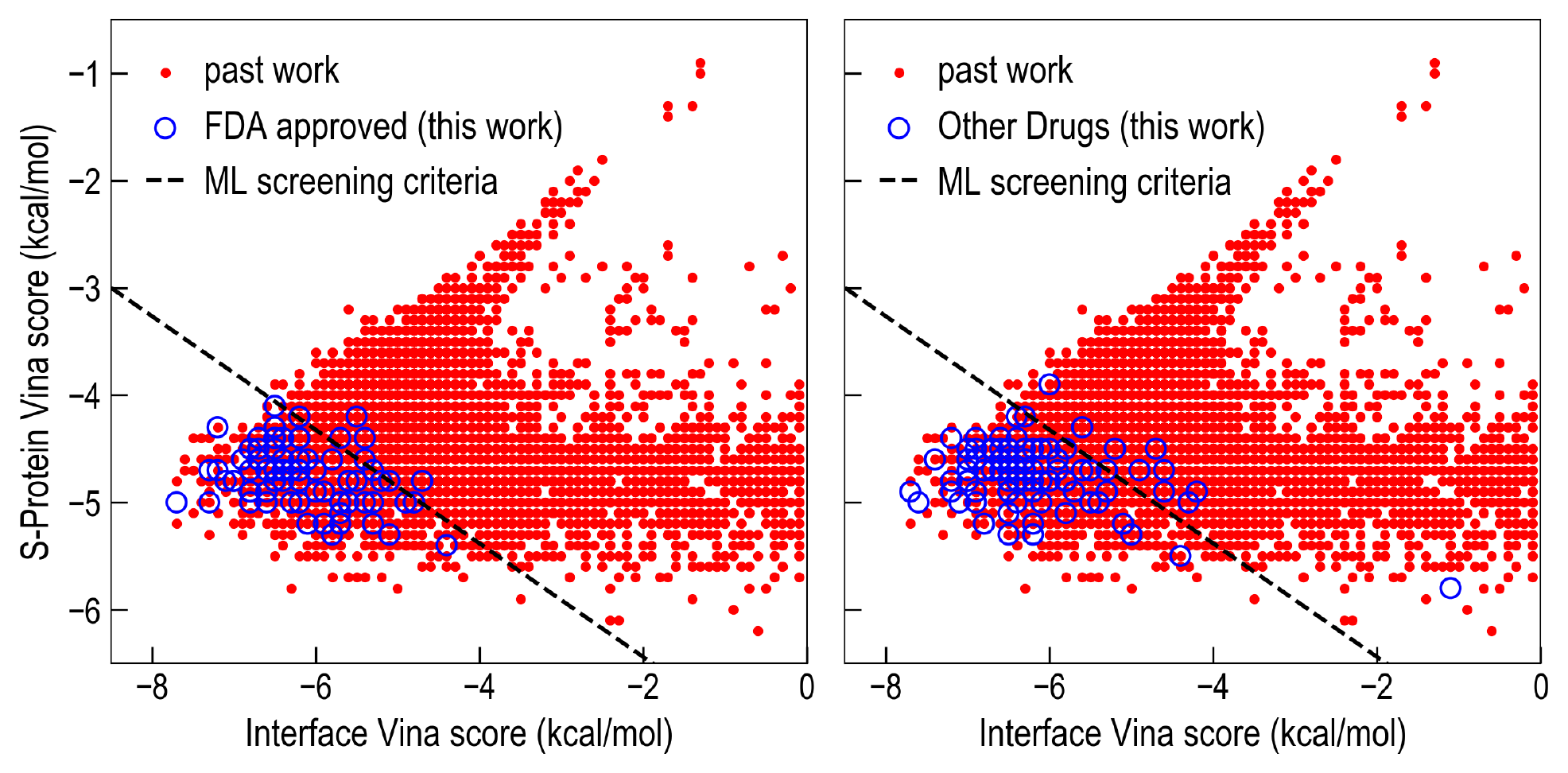}
	\caption{Vina scores for the isolated S-protein and interface between ACE2 receptor and S-protein using docking calculations. The 187 candidates selected using ML are shown in blue, and are concentrated within region of low Vina scores. For comparison, previously considered candidates from an exhaustive past work are also represented in red.}
	\label{fig:figure3}
\end{figure}

Next, ensemble docking simulations were performed for the selected 187 drug candidates, with results presented in Figure \ref{fig:figure3}. For comparison, results from the Smith dataset are also included. The purpose of these simulations was three-fold. First, a more accurate estimate of the Vina scores was obtained from these high-fidelity computations for the identified promising candidates; second, they provided new data points for further validation of the ML models, and lastly, for the 29 common candidate ligands (common to our top list and that of Smith), they help us validate our simulations against the simulations performed in the Smith paper\cite{smith2020repurposing}. From Figure \ref{fig:figure3} its evident that the ML models indeed helped us screen candidates with favorable Vina scores; almost all screened candidates can be seen to be below the ML screening criteria line (dashed line), while only 12 of the identified 187 candidates were found to have Vina scores greater than 0 and did not show any binding affinity to the S-protein:ACE2 interface complex---such cases have relatively much higher Vina score ($>$10) and are excluded from the plots for better readability. Thus, 175 of the 187 screened candidates were indeed favorable. In comparison, Smith et al. have to do expensive docking simulations for a large set of candidates, with many falling outside the screening boundary. This not only captures the efficiency of the procedure adopted here, i.e., the use of cheap surrogate models for quick screening followed by expensive high-fidelity docking simulations for validation, but also provides further validation of the prediction accuracy of the developed ML models. Parity plots directly comparing the Vina score predictions from the ML models against their respective docking simulation results are also provided in SI. Example illustrations of the S-protein:ACE2 interface-ligand complex for the top candidates are also included in the SI.

\begin{figure}
	\centering 
	\includegraphics[width=1\textwidth]{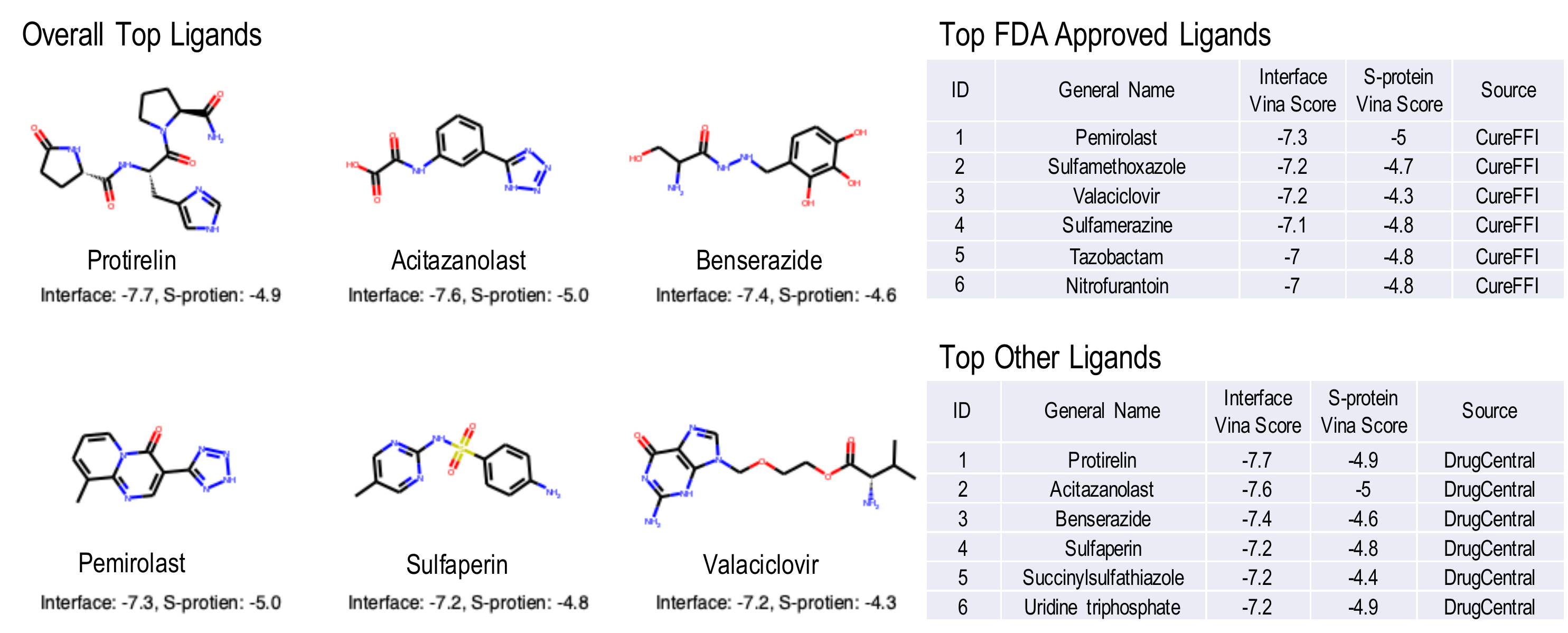}
	\caption{Top candidates identified from this work along with their Vina scores for the S-protein:ACE2 interface (labeled, interface) and the S-protein systems using the ensemble docking simulations.}
	\label{fig:figure4}
\end{figure}

More importantly, our trained ML model predicts several ligands (including several FDA approved active ingredients) with favorable Vina scores. The top 6 among the 187 candidates are presented in Figure \ref{fig:figure4} while the complete list of 187 candidates, along with their respective Vina scores, are provided in SI. The top FDA approved ligand candidates include Pemirolast (INN), which is a mast cell stabilizer used as an anti-allergic drug therapy. It is marketed under the tradenames Alegysal and Alamast. Sulfamethoxazole (SMZ or SMX), another FDA approved ligand, is an antibiotic used for bacterial infections such as urinary tract infections, bronchitis, and prostatitis. Valaciclovir is another top candidate identified from our screening and is an anti-viral drug used to treat herpes virus infections, including shingles, cold sores, genital herpes and chickenpox. Sulfanilamide is used typically as an antibacterial agent to treat bronchitis, prostatitis and urinary tract infections. Tzaobactum is another FDA approved antibiotic and is typically combined with piperacillin to treat anti-bacterial infections such as cellulitis, diabetic foot infections, appendicitis, and postpartum endometritis infection. Nitrofurantoin is also an antibiotic and used to treat urinary tract infections.

Amongst the non-FDA approved ligands, we find that the top candidate is Protirelin which is a synthetic analogue of the endogenous peptide thyrotropin-releasing hormone (TRH). Benserazide (also called Serazide) is another top ligand and is a peripherally acting aromatic L-amino acid decarboxylase or DOPA decarboxylase inhibitor that is used for Parkinson’s disease. Other top candidates include Sulfaperin (or sulfaperine), which is a sulfonamide antibacterial agent, and Succinylsulfathiazole which  is a sulfonamide used for intestinal bacteriostatic agent. Interestingly, one of the top candidates to emerge from our screening is uridine triphosphate (UTP), which is a nucleotide tri-phosphate and source of energy or an activator of substrates in metabolic reactions.

\subsection{Additional Screening Criteria: Lipinski/Pfizer Rules}
\begin{figure}
	\centering 
	\includegraphics[width=0.8\textwidth]{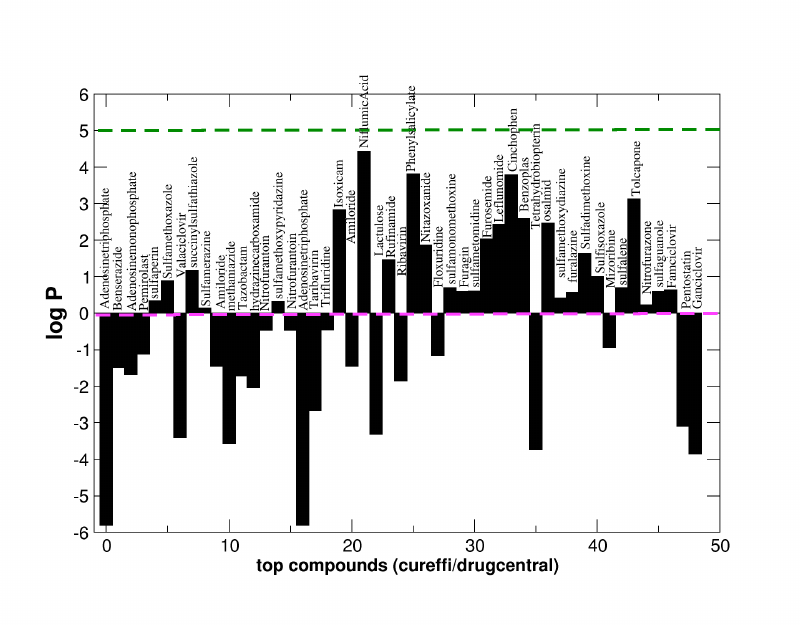}
	\caption{1-octanol/water partition coefficient (log P) of the top candidates. These values are obtained from \url{www.chemspider.com}. The green dashed line indicates a value for log P of $5$. Most of the screened top candidates have log P $<$5.}
	\label{fig:lipinski}
\end{figure}

In addition to the binding energies, one can also use other thermodynamic criteria to further screen the ligand candidates. For instance, although the binding energy of a ligand is the primary criterion and related to the binding affinity measured from the AutoDock simulations based on free energy, other metrics developed by Lipinski and co-workers \cite{lipinski1997experimental, lipinski2004lead} also have certain implications on the efficacy of the drug, and could be used to for further screening of the identified candidates. A ligand is most likely to have poor absorption when its n-octanol/water partition coefficient (log P) is $>$5, its molecular weight (MW) is $>$500, the number of H-bond donors is $>$5 and the number of H-bond acceptors is $>$10. 
Figure \ref{fig:lipinski} shows the log P of the top 50 candidates identified (based on lowest value of Vina scores) from the CureFFI and DrugCentral databases. Most of the top 50 ligands can be seen to have log P $<$5, which is consistent with Lipinski rules of five. Further, the molecular weights of the compounds are lower than 500 Da, as provided in SI along with other properties, such as Henry's constant, and number of hydrogen bond acceptors and donors.

Henry's constant (or log H) is another important property, measuring the solubility of the compound in water. For a drug to be up-taken by the cellular membrane,
it is desirable for the drug to be soluble in water.
The more negative the Henry's constant the more soluble is the drug in aqueous phase. However, a balance between desirable partitioning between the membrane and aqueous phase is generally sought. Thus, as presented in Table \ref{tab:lipinski}, the identified top candidates continue to satisfy all of the above additional criteria. Importantly, we note that more such constraints can be introduced in future work to further screen desirable candidate ligands. For instance, molecules with log P below 0 are known to have high affinity towards the aqueous media and are poorly absorbed by the lipid bilayer of the cellular membranes. Many of the top candidates can be seen to fall under this category, for e.g., Amiloride which has a log P of  $-1.45$ is water friendly and used as water-pills/diuretic. Nitrofurantoin with log P of $-0.47$ is used to treat urinary tract infections.

\begin{table}
	\centering
	\caption{n-octanol/water partition coefficient (log P), Henry's constant (log H), average molecular weight, and number of hydrogen bond donors and acceptors for the top ligands identified in this work. These values were obtained from \url{www.chemspider.com}}
	\begin{tabular}{cccccc}
		\hline
		\hline

FDA Approved Ligands &   log P    &    log H  & MW (Da) & \# of H-bond donors & \# of H-bond acceptors  \\
\hline
Pemirolast          & -1.12 &  -12.313 & 228.21 & 1 & 7 \\
Sulfamethoxazole    &  0.89 &  -10.408 & 253.278 & 3 & 6\\
Valaciclovir        & -3.41 &  -17.578 & 324.336 &	5 &	10\\
Sulfamerazine       & 0.14  &  -8.145  & 264.304 &	3 &	3\\
Tazobactam          & -1.72 & -14.714 & 300.291 &	1 &	9 \\

\hline
Other Ligands  & log P & log H  & MW (Da) & \# of H-bond donors & \# of H-bond acceptors \\
\hline

Proterelin      &  -2.46   & -22.799   &  362.384 & 5& 10\\
Acitazanolast   &  -1.95   & -16.014   & 233.184 &3 & 8\\
Sulfaperin      &  0.34   &  -8.145    & 264.304 &3 &6\\
Benserazide     &  -1.49   &    -28.420 & 257.243& 8&8\\
SuccinylSulfathiozole  &   1.18   &-19.117 & 355.389&3 &8\\
Uridine triphosphate    &   -4.09  &-38.070 &484.141 & 7&17\\

\hline
\hline
	\end{tabular}
	\label{tab:lipinski}
\end{table}

\subsection{Learned Chemical knowledge from ML model}
Beyond serving as a more computationally efficient alternative to drug docking simulations, learnt ML models can also be utilized to mine important chemical trends and extract simple chemical rules from the data. In this regard, the developed RF models can be used to identify important features/descriptors utilized in this work. In RF, the relative importance of a feature can be defined using the the relative rank (or depth) of that feature when used as a decision node in a tree, since features used at the top of a tree contribute towards the final prediction for a larger fraction of the input samples. Based on this philosophy, we provide a list of top 20 features that were found to be most relevant for the S-protein and the S-protein:ACE2 interface models in SI. Importantly, we found that the $^2\chi_{n}$ score of a molecule is very well (with Pearson correlation coefficient, R$^2$ = -0.67) correlated with its S-protein Vina score; higher the $^2\chi_{n}$ score, lower is the Vina score of the molecule:S-protein complex \cite{hall1991molecular, rdkit}. A variety of molecular quantum numbers (MQNs) were also found be highly relevant---those that captured the number of 5 or 6 membered rings, the topological surface area, cyclic trivalent and tetravalent nodes, nodes and edges shared by more than 2 rings. Count of aliphatic rings was also among important descriptors.

\subsection{ML based screening of non-FDA approved biomolecules}

\begin{figure}
	\centering 
	\includegraphics[width=1\textwidth]{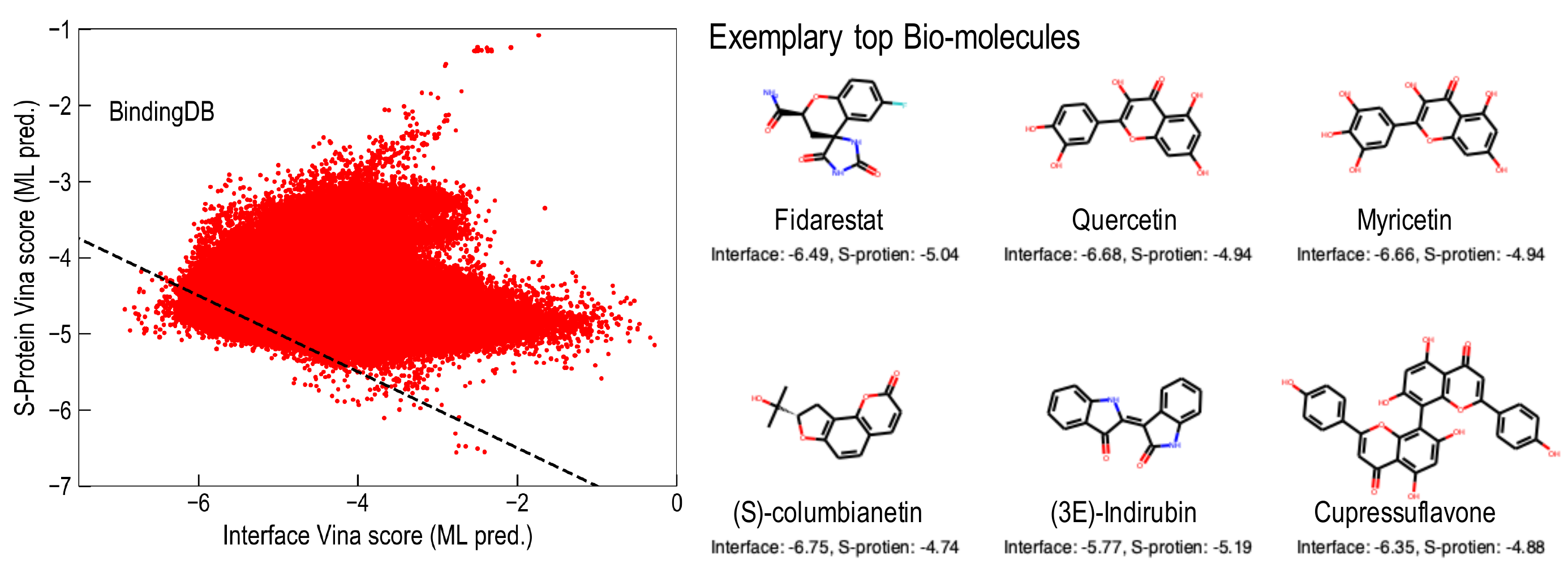}
	\caption{Vina scores predictions for the isolated S-protein and S-protein:ACE2 receptor complex for all the molecules in BindingDB dataset using ML models. Over 19,000 molecules were found to satisfy the chosen screening criteria, shown using the dashed line in the plot.}
	\label{fig:figure5}
\end{figure}

Having validated the ML models, we significantly expanded the search space of candidate molecules and made predictions for roughly 1 million molecules in the BindingDB dataset, with the Vina score predictions presented in Figure \ref{fig:figure5}(a). Nearly 19,000 molecules were found to satisfy the earlier chosen screening criteria (see SI for the complete list), and a few exemplary cases are illustrated in Figure \ref{fig:figure5}(b). These results clearly demonstrate the power and efficiency of using surrogate models for preliminary screening. For instance, the docking simulations for the identified 187 candidate active ingredients were completed in a period of around 2 days. In contrast, Vina score predictions from the ML model for the entire BindingDB dataset were obtained within a day using similar computational resources, including the time required for fingerprinting and making the model predictions. Evidently, our ML strategy is efficiently able to screen millions of candidate biomolecules and make useful suggestions to aid the decision making process for expert biologists and medical professionals. More robust high-fidelity computations, followed by synthesis and trial experiments should be performed to confirm the validity of these selected molecules.   

Amongst the screened non-FDA biomolecules, the top candidates include Fidarestat (SNK-860) which is an aldose reductase inhibitor and is under investigation for treatment of diabetic neuropathy. Quercetin is a plant flavonol from the flavonoid group of polyphenols, which also displayed high Vina scores amongst screened candidates. Other top candidates include Myricetin which is a member of the flavonoid class of polyphenolic compounds, with antioxidant properties, S-columbianetin which is used as anti-inflammatory, Indirubin that has anti-inflammatory and anti-angiogenesis properties in vitro and Cupressuflavone with anti-inflammatory and analgesic properties. 

\section{Conclusions}
In conclusion, we present an efficient virtual screening strategy to identify ligands that can potentially limit and/or disrupt the host-virus interactions. Our hypothesis is that ligands that bind strongly to the isolated SARS-CoV-2 S-protein at its host(human) receptor region and to the Sprotein-human ACE2 interface complex are likely to be most effective. Our high-throughput screening strategy is based on using a combination of machine learning (ML) and high-fidelity docking simulations to identify candidates that display such high binding affinities. We first train ML on results of computationally expensive simulations, and subsequently use the validated ML to search a much larger chemical space ($\sim$1000's of FDA approved ligands and subsequently $\sim$millions of biomolecules). We down-select based on ML predicted Vina scores, and finally mine chemical guidelines to accelerate the therapeutic cure of diseases.

Two random forest models were trained to quickly predict Vina scores of molecules with isolated S-protein and Sprotein:human ACE2 interface complex, using dataset from the past work. To train the ML models, a comprehensive set of chemical features was compiled, based on our past experience on fingerprinting organic materials, to capture geometric and chemical information about ligands at multiple length-scales. The ML models were first used to screen 187 ligands from two drug datasets (CureFFI and DrugCentral), 175 of which were indeed found to bind strongly to the isolated S-protein and to the Sprotein-human ACE2 interface complex using the expensive AutoDock simulations. This not only validates the accuracy of the ML models developed here, but also helped to identify 75 promising FDA approved ligands. Many of the identified top ligands were also found to satisfy Lipinski's rule of five. With the ML models validated, we used them to screen $\sim$19,000 candidates from a large dataset of bio-molecules ($\sim$ 1 million) from the BindingDB dataset. A rank-ordered list of promising candidates from the different datasets are provided for further theoretical or experimental validation.

\section{Acknowledgements}
This work was performed at the Center for Nanoscale Materials, a U.S. Department of Energy Office of Science User Facility, and supported by the U.S. Department of Energy, Office of Science, under Contract No. DE-AC02-06CH11357. We thank Naina Zachariah for useful discussions and for reviewing the manuscript. SKRS thanks UIC Start-up faculty grant for supporting this work.

\bibliographystyle{unsrt}

\newpage
\section{TOC Graphic}
\begin{figure}[!htb]
	\centering 
	\includegraphics[width=1\textwidth]{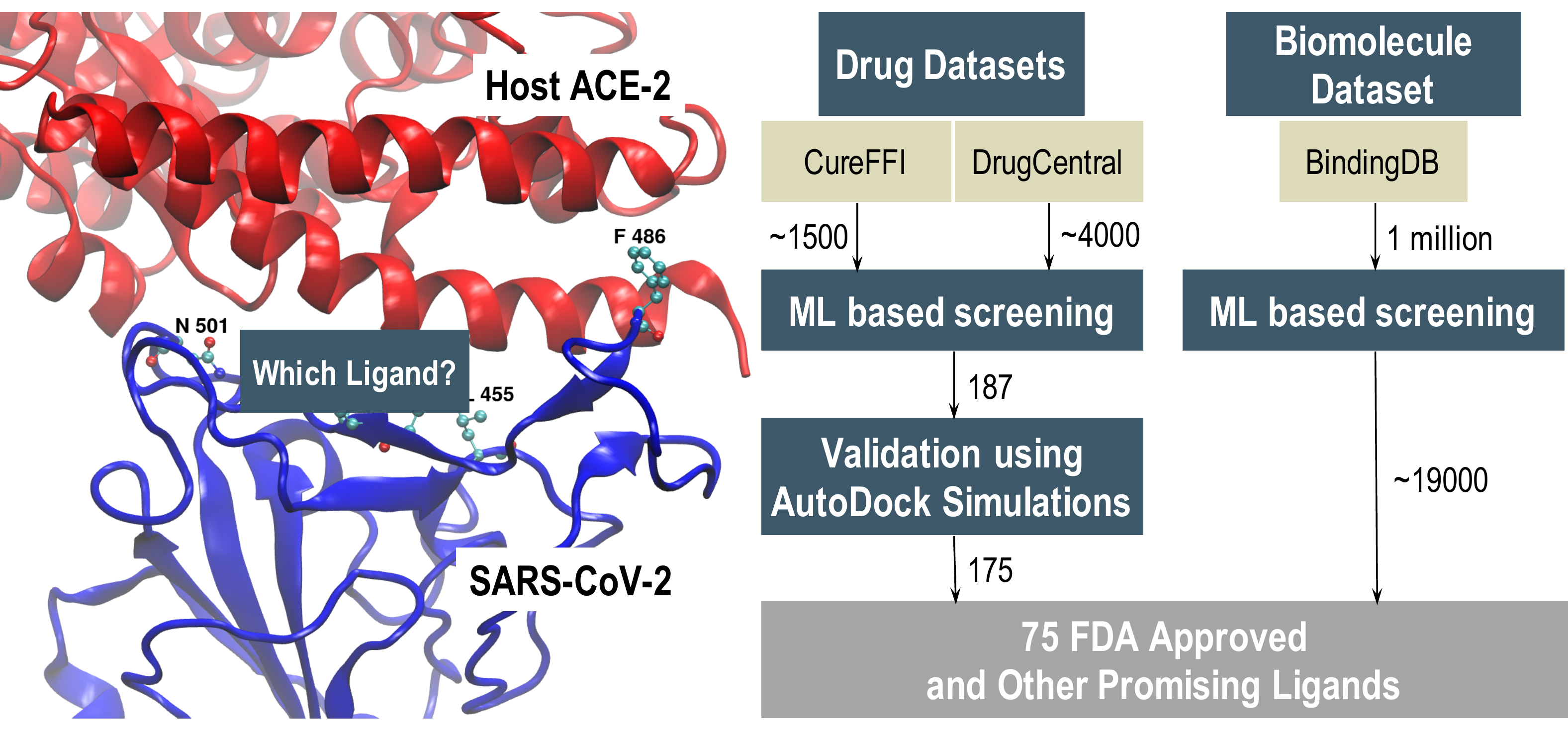}
	\caption{TOC Graphic}
	\label{fig:TOC}
\end{figure}

\end{document}